\documentclass{article}
\usepackage[dutch, english]{babel}
\usepackage{amssymb}
\usepackage{amsmath}
\usepackage{mathrsfs}
\usepackage{latexsym}
\usepackage{longtable}
\usepackage{multirow}
\usepackage{epsf}
\usepackage{color}
\usepackage{graphicx}
\usepackage{float}
\usepackage{enumerate}
\usepackage{txfonts}
\usepackage{natbib}

\usepackage{asp2010}

\begin{document}

\aspvoltitle{Circumstellar Dynamics at High Resolution}

\title{The star formation history of RCW~36}

%\authorrunning{L.E. Ellerbroek et al.}
%\titlerunning{History of RCW 36}

\author{L.~E.~Ellerbroek\altaffilmark{1}, L.~Kaper\altaffilmark{1}, A.~Bik\altaffilmark{2}, K.~M.~Maaskant\altaffilmark{1}, L.~Podio\altaffilmark{3,4}}
\altaffiltext{1}{Sterrenkundig Instituut ``Anton Pannekoek", University of Amsterdam, Science Park 904, P.O. Box 94249, 1090 GE Amsterdam, The Netherlands}
\altaffiltext{2}{Max-Planck-Institut f\"{u}r Astronomie, K\"{o}nigstuhl 17, 69117 Heidelberg, Germany}
\altaffiltext{3}{Kapteyn Astronomical Institute, Landleven 12, 9747 AD Groningen, The Netherlands}
\altaffiltext{4}{Laboratoire d'Astrophysique de l'Observatoire de Grenoble, UMR 5521 du CNRS, 38041 Grenoble Cedex, France}

\begin{abstract}
Recent studies of massive-star forming regions indicate that they can contain multiple generations of young stars. These observations suggest that star formation in these regions is sequential and/or triggered by a previous generation of (massive) stars. Here we present new observations of the star forming region RCW~36 in the Vela Molecular Ridge, hosting a young cluster of massive stars embedded in a molecular cloud complex. In the periphery of the cluster several young stellar objects (YSOs) are detected which produce bipolar jets (HH~1042 and HH~1043) demonstrating that these objects are still actively accreting. The VLT/X-shooter spectrum of the jet structure of HH~1042 provides detailed information on the physical conditions and kinematical properties of the jet plasma. From this information the YSO's accretion history can be derived. Combining the photometric and spectroscopic observations of RCW~36 gives insight into the formation process of individual stars and the star formation history of this young massive-star forming region.
\end{abstract}

\section{Introduction}

Important questions in the hitherto poorly understood process of (massive) star formation are: (i) How are the most massive stars formed? (ii) Which processes trigger star formation? We approach these questions by studying young stars and their environment during or directly after the time of formation. Near-infrared imaging surveys provide a census of the embedded young stellar populations. Follow-up near-infrared spectroscopy provides a window on the photospheres of young massive stars, as the extinction is orders of magnitude less than at UV and optical wavelengths, and thermal dust emission has not set in. This then results in accurate spectral classification and allows for a detailed investigation of the signatures of star formation (e.g. disks and jets). 

Massive stars ($M> 8$ M$_\odot$) are rare, their formation timescale is short and their birth places are obscured by dust, leading to high extinction (A$_{V} \sim 10-100$~mag). These factors make it difficult to observe massive stars in formation; see e.g. \citet{Zinnecker2007} for a review. Massive stars are expected to form in a similar way to low-mass stars, with the difference that the strong UV radiation field of the contracting massive star will hinder the accretion process. The first detectable phase is thought to be the ultra-compact H~{\sc ii} region, when the young massive star starts to ionize the surrounding medium. Whether or not the central star(s) embedded in an UCH{\sc ii} are still contracting to the main sequence and carry the signatures of the formation process remains to be demonstrated. Theoretical models by \citet{Hosokawa2010} predict that massive stars can form through disk accretion, though require high accretion rates ($\dot{M} > 10^{-4}$~M$_\odot$~yr$^{-1}$). During this rapid accretion process, the radius of the star will increase dramatically due to the high entropy deposited on the stellar surface. At later stages, the protostar begins to contract towards the main sequence. Thus, \citet{Hosokawa2010} predict the existence of bloated high-mass protostellar objects, which lack detectable H~{\sc ii} regions. An example of such an object is B275 in M17, an 8 M$_\odot$ pre-main sequence (PMS) star with the dimensions of a giant \citep{Ochsendorf2011}.

Most massive stars form in clusters \citep[e.g.,][]{Lada2003, Zinnecker2007}. Both during and after their formation, they will have impact on their surroundings. In some well-studied massive star-forming regions, the massive stars are centrally located and appear to have formed later than an older generation of intermediate-mass stars \citep[e.g.,] []{Feigelson2008, Bik2010, Maaskant2011}. 
Sequential star formation is not a new phenomenon \citep[see e.g.][]{Blaauw1964, Blaauw1991}. An unanswered question is whether massive stars play an important role in this sequence; whether a new generation is triggered by their emergent radiation field, stellar winds and supernovae, or perhaps that massive stars are the last to form in a cluster. Several scenarios exist for triggered star formation; massive stars are expected to play a crucial role in this process \citep{Andre2009}. 

\begin{figure*}[!t] 
   \centering
\includegraphics[width=0.5\textwidth]{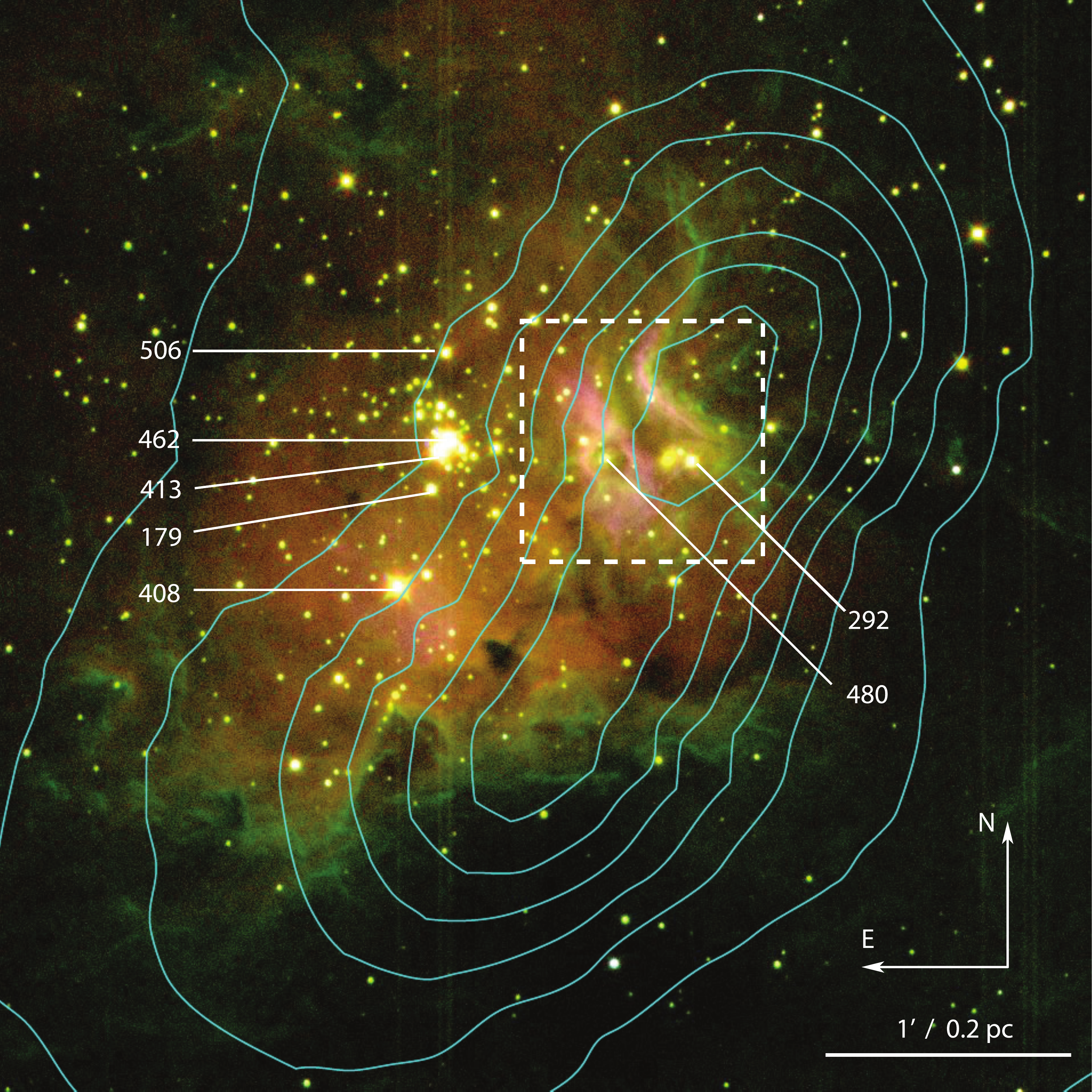} 
   \includegraphics[width=0.48\textwidth]{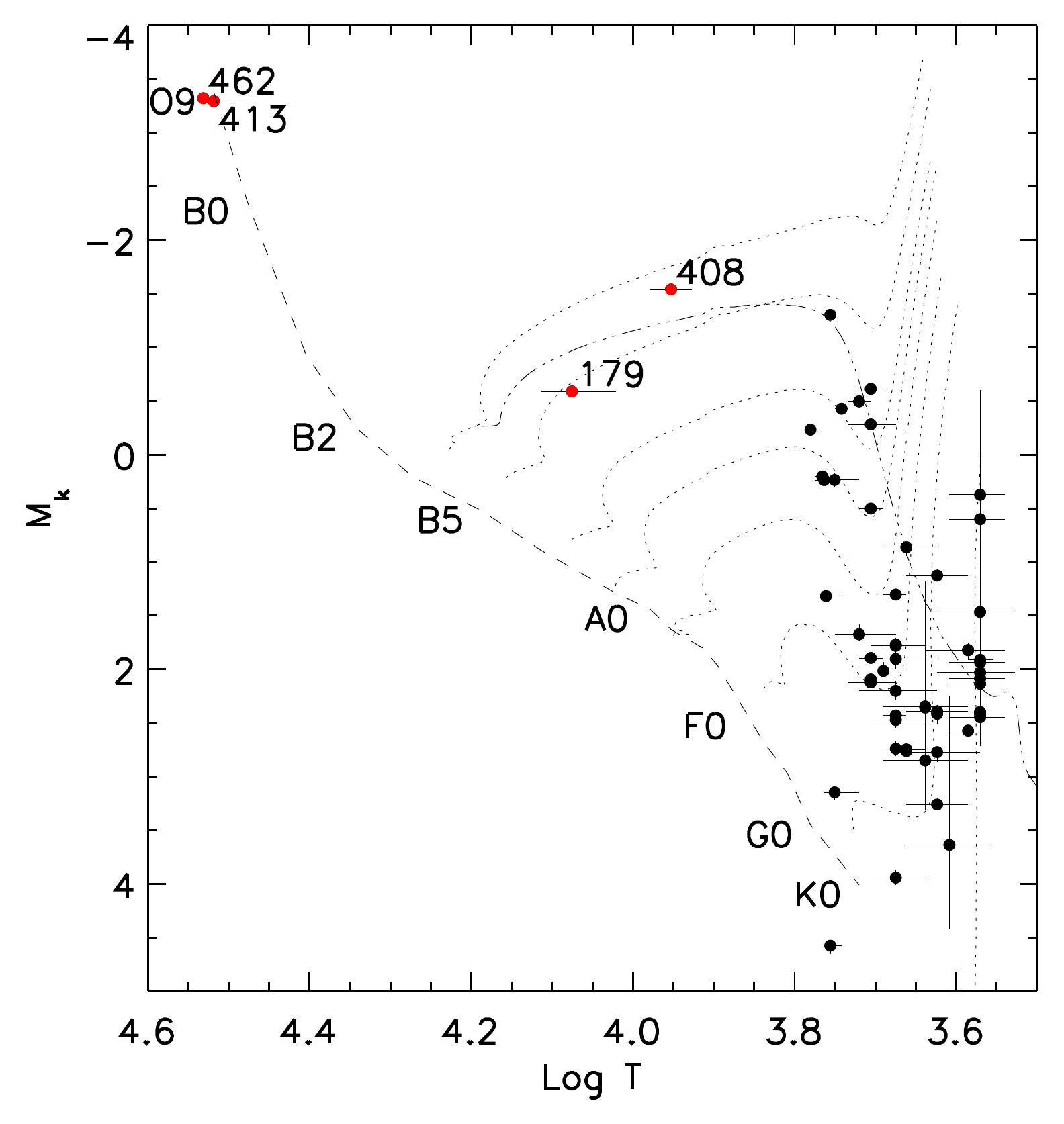} 
   \caption{\textit{Left:} NTT/SOFI narrowband image of RCW 36 (\textit{red:} Br$\gamma$, \textit{green:} H$_2$, \textit{blue:} Pa$\beta$). Overplotted contours are BLAST 500 $\mu$m flux, tracing the molecular cloud \citep{Netterfield2009}. X-shooter targets are labeled. The region within the dashed square is displayed in Fig.~\ref{fig:linemap}. \textit{Right:} Hertzsprung-Russell diagram for RCW 36, with an adopted distance of 0.7~kpc. Classified stars which were also observed with X-shooter are plotted with red symbols. Dotted lines are pre-main sequence (PMS) tracks (5, 4, 3, 2.5, 2, 1.5, 1 and 0.5 M$_\odot$, from top to bottom) and the dash-dotted line is the 1~Myr PMS isochrone \citep{Siess2000}. The intrinsic near-infrared excess of 408 was subtracted from its K-band flux.}
   \label{fig:map}
\end{figure*}

In order to understand this process in more detail, a complete census of the outcome of star formation is required. Regions located relatively nearby are best suited for this purpose. However, nearby star-forming regions contain few or no massive stars (Taurus-Auriga) or their star formation process is already in an advanced stage (Orion), while most massive-star forming regions are distant ($\ge$ 2 kpc, e.g. M17, \citealt{Hanson1997}; 30 Dor, \citealt{Walborn1997}).

In this paper, we present a near-infrared imaging and spectroscopy survey of the massive star-forming region RCW 36. The optical to near-infared spectrograph X-shooter, mounted on the ESO \textit{Very Large Telescope}, is used to probe the youngest and the most massive objects in this cluster in order to obtain information about their photospheres and circumstellar material. The region contains two jet sources; their spectra contain a fossil record of their recent outflow history. In section~\ref{sec:sfr} the star forming region, along with its infrared-brightest stars, is introduced. Section~\ref{sec:hhobjects} then elaborates on the two jet sources (HH~1042 and HH~1043). The physical conditions of HH~1042, along with a simulation of its kinematics with a simple ballistic model, are presented in section~\ref{sec:model}. The results are put in the context of sequential star formation in section~\ref{sec:discussion}.

\begin{figure*}[!t] 
   \centering
   \includegraphics[width=\textwidth]{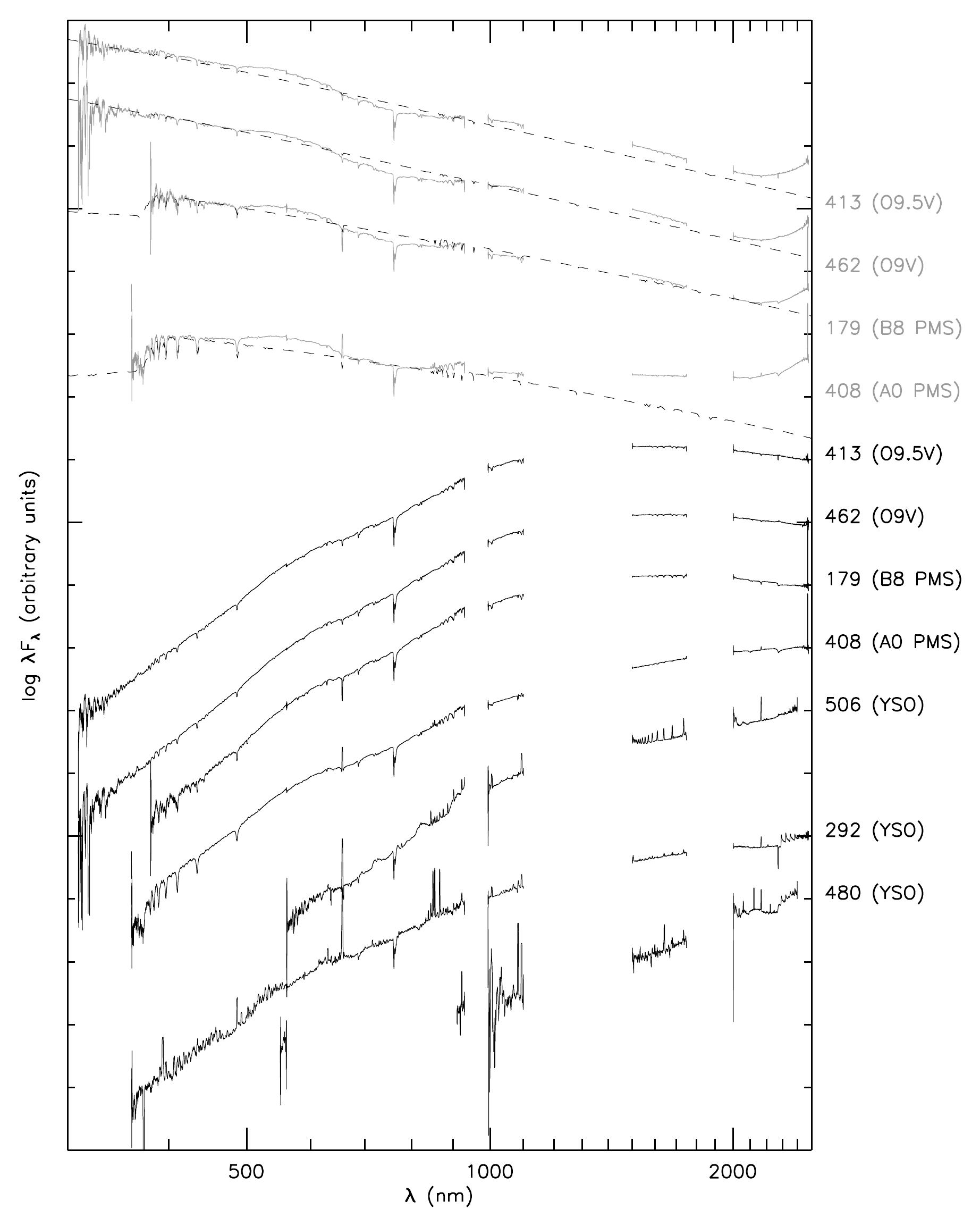} 
   \caption{Spectral energy distributions of the X-shooter sample of RCW 36. The spectra are normalized to their mean K-band flux. Overplotted~in grey are the dereddened spectra of the four classified stars. Dashed lines are Kurucz models corresponding to the spectral type, scaled on the dereddened flux. Note the intrinsic infrared excess redward of 1000 nm in 408.}
   \label{fig:seds}
\end{figure*}
\clearpage

\begin{figure*}[!t] 
   \centering
   \includegraphics[width=\textwidth]{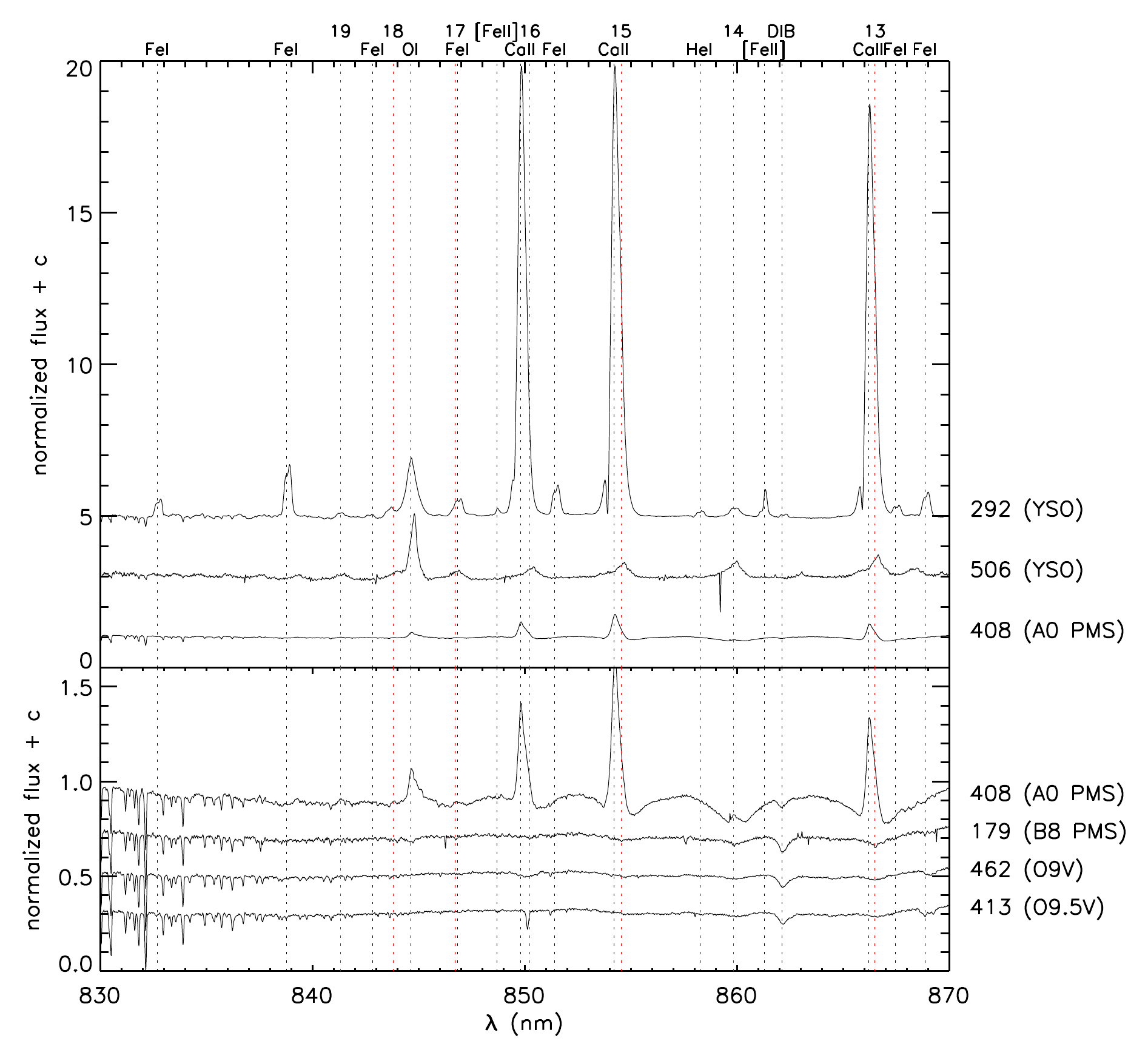} 
   \caption{Normalized spectra in the Ca~{\sc ii} triplet region of the stars in RCW~36 observed with VLT/X-shooter. Note the absence of the diffuse interstellar band in the younger objects.}
   \label{fig:normspec}
\end{figure*}

\section{The star-forming region RCW 36}
\label{sec:sfr}

A good opportunity to study the sequence of star formation is presented by the cluster RCW~36, part of the Vela~C molecular complex, at a distance of $0.7 \pm 0.2$~kpc \citep{Liseau1992}. Photometric data from NTT/SOFI and \textit{Spitzer}/IRAC, integral field spectroscopy from VLT/SINFONI, spectroscopy from VLT/X-shooter \citep{Ellerbroek2012a} and \textit{Herschel}/PACS and SPIRE observations \citep{Hill2011, Giannini2012} reveal a rich stellar population and cloud environment (Fig.~\ref{fig:map}, left). The center of the star forming region hosts a young cluster including a number of OB stars \citep{Bik2005}, while also numerous Young Stellar Objects (YSOs) are identified, two of which have strong CO ($v=2-0$) emission produced by a dense disk \citep{BikThi2004, Wheelwright2010}. Both YSOs produce extensive jets with atomic gas velocities of several hundred km~s$^{-1}$ \citep{Ellerbroek2011}, indicating that they are still actively accreting. The region suffers from a high optical exinction due to the local presence of interstellar dust. By dereddening all sources to the $JHK$ colors of classical T Tauri stars \citep{Meyer1997} we find an average value of A$_V = 10.5$~mag, somewhat higher than A$_V = 8.1$~mag found by \citet{Baba2004} by the same method. Possible explanations for this difference are the better observing conditions and higher completeness (up to $J$=19) of the dataset used in the current study.

The spectral type and luminosity class of 58 sources were determined by their $H$- and $K$-band spectra. Four early-type stars were observed with VLT/X-shooter (300 -- 2500 nm) and classified by their optical spectra. The results of this classification are contained in the Hertzsprung-Russell diagram (HRD, Fig.~\ref{fig:map}, right). A pre-main sequence population is scattered around the 1~Myr isochrone, while the two central OB stars (413, O9.5V and 462, O9V) are already on the main sequence. The distribution of  the stellar masses according to the evolutionary tracks is consistent with the Salpeter initial mass function for stars in the Galaxy \citep{Salpeter1955}.

In Figs.~\ref{fig:seds}~and~\ref{fig:normspec} spectral energy distributions (SED) and normalized spectra are displayed of the most significant sources in the cluster. The two O~stars 413 and 462 are almost featureless, save for diffuse interstellar bands (DIBs). These DIBs are also seen in the other two classified stars, while they are absent in the YSO spectra. 

The two most massive PMS stars are 179 (B8) and 408 (A0), on evolutionary tracks of 4 and 5~M$_\odot$, respectively. The \textit{Spitzer} observations of 179 show no infrared excess at all, nor does it exhibit any emission lines. We conclude it is a bloated PMS star, not yet on the main sequence. Object 408 has an infrared excess redward of 1000~nm (see Fig.~\ref{fig:seds}) with colors characteristic of a Lada class II source \citep{Lada1987}. It also exhibits H~{\sc i} emission lines superposed on broad absorption lines. Towards lower transitions the lines become optically thick and have P-Cygni profiles indicative of a wind origin. The Ca~{\sc ii} triplet profiles have an extended red wing, also a wind diagnostic.

\begin{figure}[!t] 
   \centering
   \includegraphics[width=0.58\textwidth]{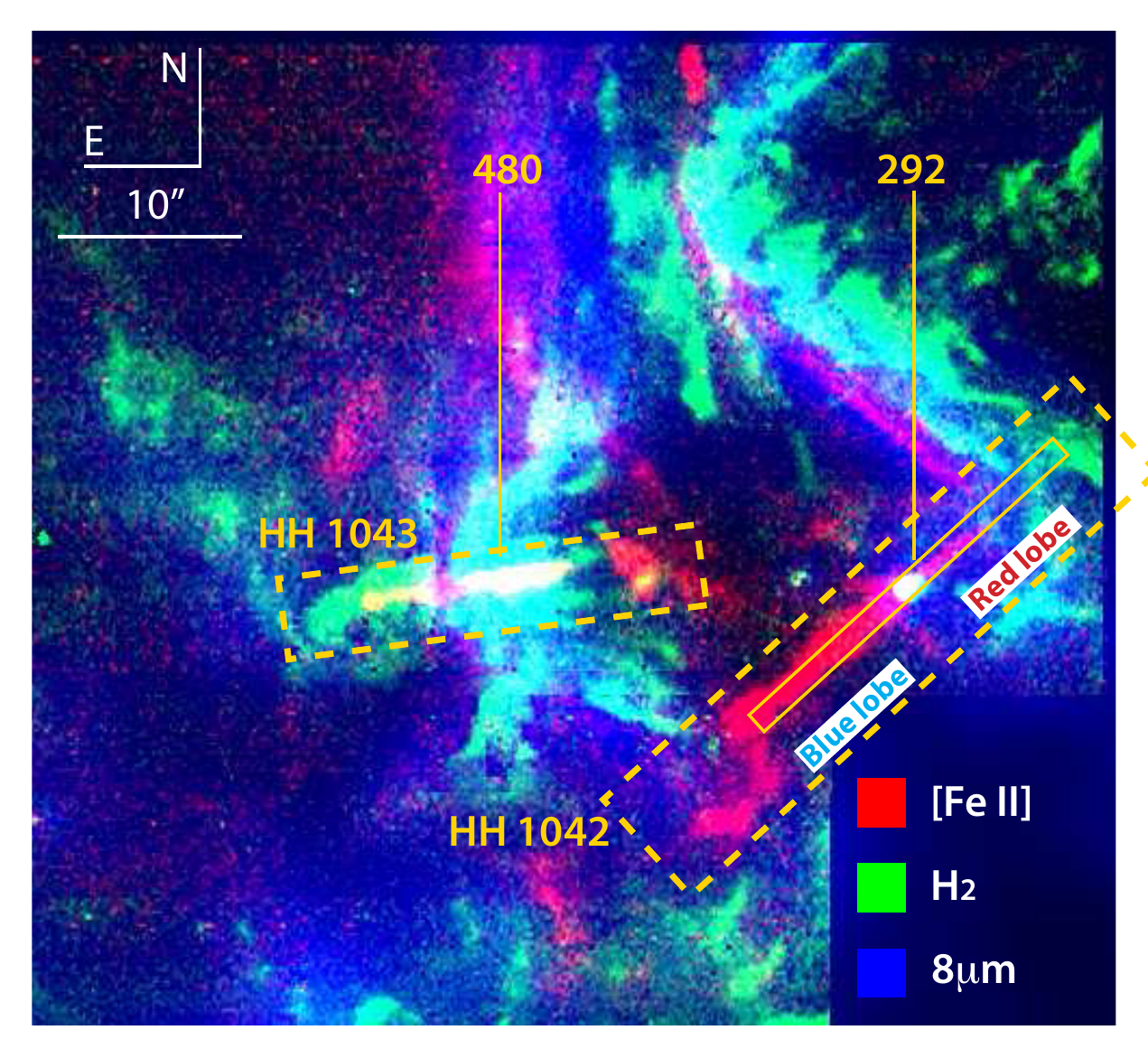} 
   \includegraphics[width=0.41\textwidth]{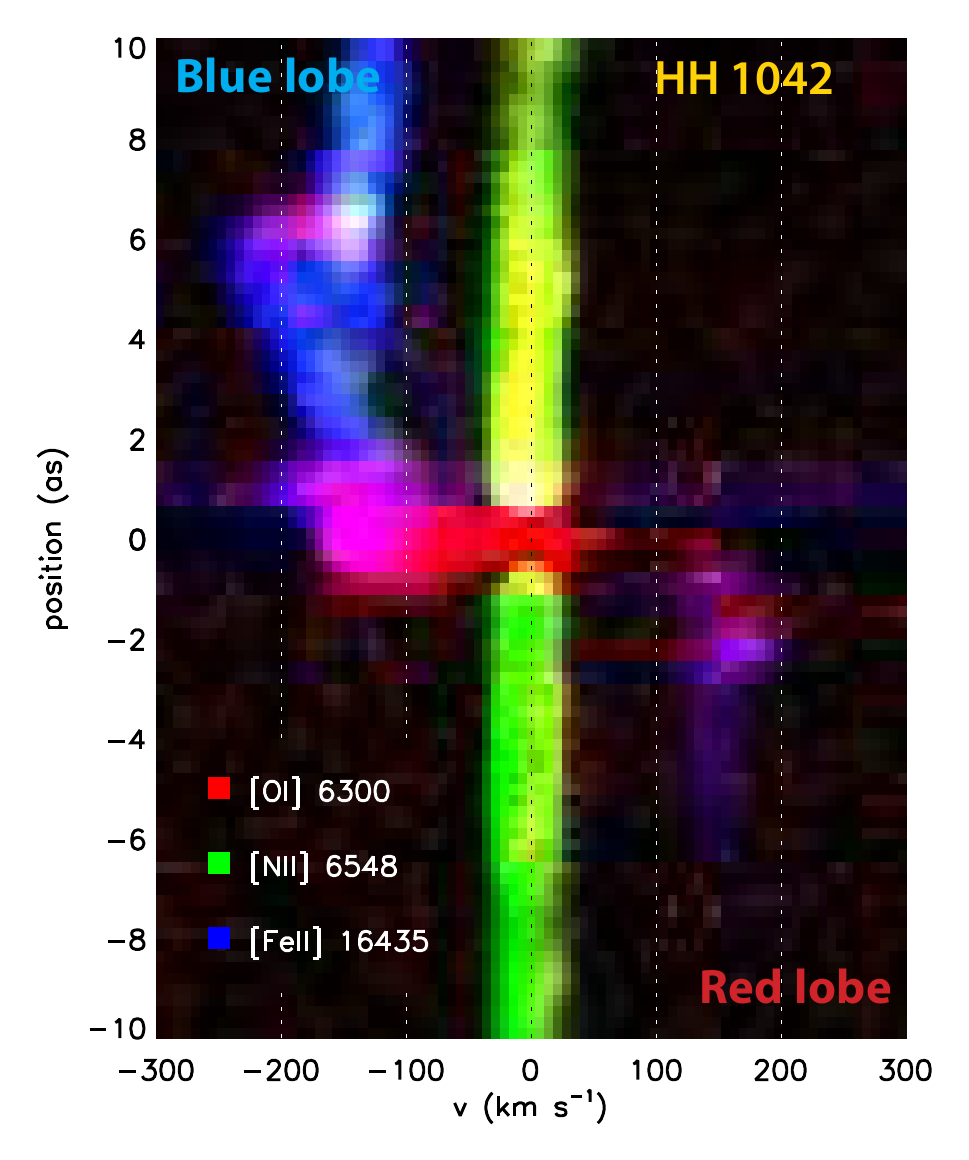} 
      \caption{\textit{Left:} Composite linemap of the dashed region indicated in Fig.~\ref{fig:map}. Colors trace the [Fe~{\sc ii}]~$\lambda$1644~nm and H$_2$ $\lambda$2121~nm lines as observed with VLT/SINFONI, and the Spitzer/IRAC 8~$\mu$m continuum. The [Fe~{\sc ii}] emission traces the jets. \textit{Right:} Composite linemap of the X-shooter spectrum of HH~1042. Colors trace three emission lines ([O~{\sc i}], [N~{\sc ii}] and [Fe~{\sc ii}]); the fluxscale is normalized to the 1$\sigma$ noise level. The central source continuum at $0''$ is subtracted using a Gaussian fit. Note the difference between the [O~{\sc i}] and [Fe~{\sc ii}] emission. The emission at zero velocity is of nebular origin.}
   \label{fig:linemap}
\end{figure}

The three other sources in Fig.~\ref{fig:seds}~and~\ref{fig:normspec} are identified as YSOs by their infrared excess and pure emission line spectra indicative of disks and outflows. Object 506 is an embedded class I source; its H~{\sc i} emission lines have extended red wings. Object 292 is an actively accreting system, its photosphere veiled by continuum emission from the accretion disk and its spectrum dominated by emission lines showing signatures of Keplerian rotation and outflow. The spectrum exhibits many H~{\sc i}, He~{\sc i} and metallic lines, and the 2290~nm CO bandheads are in emission. Also, its many forbidden lines are associated with a (disk) wind or jet, which is also visible in the VLT/SINFONI [Fe~{\sc ii}] image (HH 1042, Fig.~\ref{fig:linemap}). For an extensive discussion on the nature of object 292, see \citet{Ellerbroek2011}.

Object 480 has a spectrum (not shown in Fig.~\ref{fig:normspec}) dominated by emission lines from both the nebula and a jet. Only the Ca~{\sc ii} triplet, CO bandheads and a weak continuum redward of the H band can be related to the central object. Like 292, it powers a jet (HH~1043) visible in many H~{\sc i} and forbidden emission lines. 

From top to bottom, the spectra in Fig.~\ref{fig:normspec} show a tentative sequence of star formation, from embedded objects with disks and outflows, via PMS stars with and without disks and winds, to main sequence stars with photospheric spectra void of any emission lines. Also note the increasing strength of the DIB feature at 862~nm along this sequence. The fact that these objects are all found in the same cluster rises the question whether this is evidence for sequential or even triggered star formation. No direct evidence for binarity is found in these data. In the remainder of this paper, we focus on two individual objects, namely the jets HH 1042 and HH 1043, powered by 292 and 480, respectively.

\section{HH~1042 and HH~1043}
\label{sec:hhobjects}

Jets are a universal signature of angular momentum transport in accreting systems. The observational signature of jets emitted by YSO are the so-called Herbig Haro objects. These are ionized nebulae produced by the shocks occurring along the jet and become optically visible in the late stages of the star formation process, when the environment has been cleared by the jet itself.   Stellar jets are thought to be powered by a stellar or disk wind, carrying away material from the disk along magnetic fieldlines by the centrifugal force. The confinement of the plasma along magnetic field lines results in a collimated flow \citep{Blandford1982}.

The gas heated behind the shock fronts produce emission lines which can be used to retrieve the gas physical conditions and kinematics. For example, the analysis of the velocity variation along the flow allows us to investigate the formation of shocks. These can have multiple origins, e.g. a variable outflow rate at the base or interaction with the ambient nebula. In \S \ref{sec:physcond} we briefly present the characteristics of the two Herbig-Haro objects in RCW~36. In \S \ref{sec:model} we reconstruct the likely accretion history from the kinematics of the gas in one of these jets.

\citet{Ellerbroek2011} report the detection of two jets in RCW~36 in the [Fe~{\sc ii}]~1644~nm and Br$\gamma$ images obtained with VLT/SINFONI. They have been classified as a Herbig-Haro objects HH~1042 and HH~1043 (B. Reipurth, private communication). The left panel of Fig.~\ref{fig:linemap} shows the location of the jets in the star forming region. Note that they are located in the periphery of the cluster, along the filaments traced by Br$\gamma$ and H$_2$ emission (Fig.~\ref{fig:map}, left). A lack of mid-infrared continuum emission is detected around HH~1042, which suggests that this jet is protruding from the molecular cloud. The northwestern (redshifted, or `red') lobe fades into an ISM structure, while the southeastern (`blue') lobe seems to be colliding with the ISM. It is not clear whether the patch of [Fe~{\sc ii}] emission 8$''$ southeast of this location is related to the jet. HH~1043 is more embedded; both mid-infrared and H$_2$ emission is detected near with the jet. The western lobe terminates in a bow shock feature dominated by [Fe~{\sc ii}] emission.

Optical to near-infrared, medium-resolution (25 $< \Delta v <$ 50~km~s$^{-1}$) spectra of the two jets were obtained with VLT/X-shooter, with the slit aligned along the jet axis.. In the next sections the X-shooter spectrum of HH~1042 is presented and discussed.

\subsection{Physical and kinematical structure of the HH~1042 jet} 
\label{sec:physcond}

The VLT/X-shooter spectrum of HH 1042 shows emission in more than 90 emission lines. Line velocity variations on an arcsecond scale suggests the interaction of gas components with different velocities. Fig.~\ref{fig:linemap} (right panel) shows a composite position velocity (PV) diagram of three emission lines which trace different gas components in the flow. Note that the emission lines in the red lobe are suffering from extinction due to the molecular cloud.

The [N~{\sc ii}] line, tracing a medium with a high degree of ionization, dominates at 0~km~s$^{-1}$. This component is likely background emission from the H~{\sc ii} region and is unrelated to the jet. Fe is strongly depleted in the ISM and released in gas-phase at the jet launch and/or in the shocks occurring along the jet. Hence, the [Fe~{\sc ii}] line is exclusively detected in the jet and reveals its kinematic structure. We observe a varying velocity along the flow; at $-2''$ and $+6''$ shocks form where fast material is `ploughing' into slower material. These regions are traced by [O~{\sc i}] and [N~{\sc ii}] emission. At the base of the jet, before strong ionization is produced by the shocks, we detect bright [O~{\sc i}] emission, possibly tracing the disk wind origin of the jet. Its velocity at the base coincides with the blueshifted absorption component of the Ca~{\sc ii} triplet lines (see Fig.~\ref{fig:normspec}), strongly suggesting a connection between the jet and a disk wind, although these spectral lines may trace different parts of the flow. 

\begin{figure}[!b] 
   \centering
   \includegraphics[width=\textwidth]{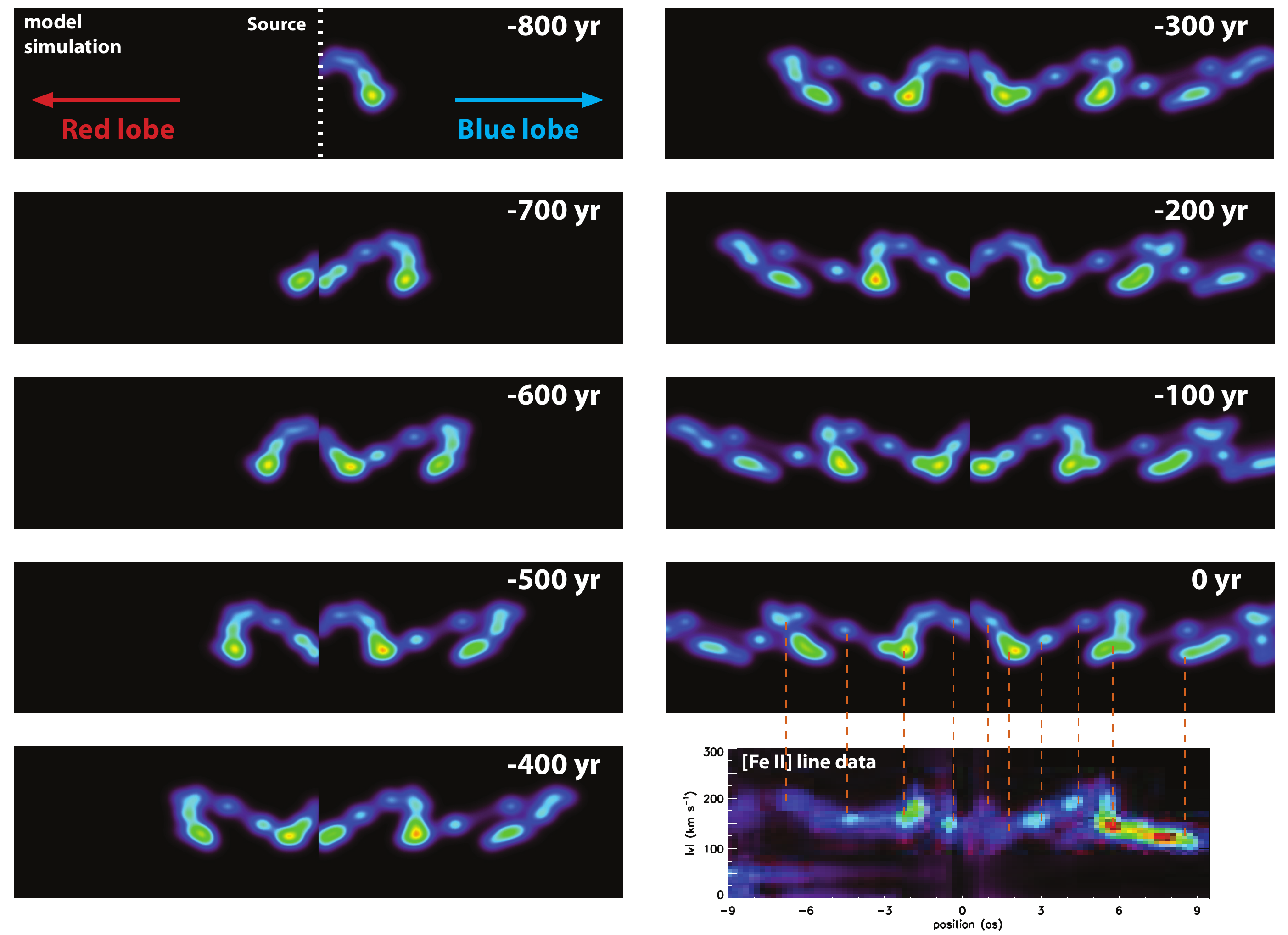} 
   \caption{Model simulation assuming a flow with parameters specified in the text. 
   The observed [Fe~{\sc ii}]~$\lambda$1644 emission position-velocity diagram (bottom-right panel) is compared with the result from the model simulation. The adopted outflow rate and its parameters are specified in the text. The dashed lines indicate the knot positions in the observed and simulated position-velocity diagrams.}
   \label{fig:jetmovie}
\end{figure}

\subsection{The accretion history of HH~1042}
\label{sec:model}

We now present the analysis of the kinematics (positions and velocities) of the brightest jet tracer in the spectrum ([Fe~{\sc ii}]~$\lambda$1644~nm). We adopt a method first used by \citet{Raga1990} to determine the outflow history of the jet. A concise model is conceived in order to simulate the flow over the past 900 years. 

We start by measuring the positions on the sky $x_{t}$ and radial velocities $v_r$ of the local emission maxima, or `knots' (see Fig.~\ref{fig:jetmovie}, lower right panel). Assuming that the jet is perpendicular to the disk, and adopting the value $i=18^\circ$ (the inclination of the jet axis with respect to the line of sight) derived by \citet{Wheelwright2010}, we can then calculate for each knot a timescale, which can be interpreted as the time at which the knot was launched, given no collisions have occured in the flow:
\begin{equation}
t_{\rm launch} = \frac{x_t}{v_r \tan i}
\end{equation}

We then fit a periodic outflow rate at the jet base to the resulting values of $v=v_r / \cos i$ and $t_{\rm launch}$:
\begin{equation}
v(t) = 170\hspace{3pt}{\rm km~s^{-1}} + 42\hspace{3pt}{\rm km~s^{-1}} \sin \left(2\pi \frac{t}{280\hspace{3pt}{\rm yr}}\right)
\end{equation}
We furthermore note that the knots in the red lobe are fit better by this function if it is offset by a phase difference of $\pi/2$. Also, an additional mode with a period of 32 yr is introduced to produce the substructure along the flow. We use this function as an input for a ballistic model, where no collisions occur along the flow, assume a constant mass loss rate, and run a simulation for $t=[-1000, 0]$~yr (Fig.~\ref{fig:jetmovie}). The global emission profile of the [Fe~{\sc ii}] line can be reproduced reasonably well.  It should be noted that the colors in the simulation images in Fig.~\ref{fig:jetmovie} trace the material density, while the observed quantity is line emission. The derived velocities and timescales, as well as the apparent phase shift between the two lobes, can put constraints on theoretical models for the jet launching mechanism. For a detailed discussion we refer to \citet{Ellerbroek2012b}.

\section{Discussion}
\label{sec:discussion}

The stellar population of RCW~36 consists of a large number of low-mass PMS stars ($<2$M$_\odot$), a couple of intermediate-mass PMS stars (2--8~M$_\odot$) and two high-mass stars ($\sim 20~$M$_\odot$) already on the main sequence. In addition to this, at least three YSOs are present,  of which the spectral type cannot be determined; based on their spectral characteristics (CO, H~{\sc i}, He~{\sc i}) their central objects are luminous intermediate- to high-mass stars \citep{Ellerbroek2011}. Two of these are associated with HH outflows, indicating that accretion is still ongoing.

The simultaneous presence of low-mass PMS stars and intermediate-mass YSOs suggests that these belong to different generations of star formation. The low- to intermediate-mass PMS stars seem to have formed about 1~Myr ago. The O~stars are close to the ZAMS, implying they are of an equal age or younger ($\lesssim 1$~Myr), if their PMS phase is indeed as rapid as suggested by observations \citep[$\sim 10^5$~yr,][]{Mottram2011}. Shortly after having formed they possibly triggered the formation of the YSOs in the periphery of the cluster. This is also suggested by the presence of a shock wave (the filament traced by Br$\gamma$ and H$_2$ emission, Fig.~\ref{fig:map}, left). The dynamical timescale associated with the distance between the O~stars and the shock wave (0.2 pc), assuming it has propagated with a typical soundspeed of a~few~km~s$^{-1}$, is of the order of 10$^4$~yr. This, and the young age of the YSOs, is consistent with the YSOs having formed after the massive stars. However, this scenario cannot be confirmed, mainly because the age of the O~stars is poorly constrained due to the uncertainty in the distance to the cluster (i.e. the precise location of the ZAMS). 

Recent studies \citep{Feigelson2008, Bik2010, Wang2011, Maaskant2011} have found similar age spreads in massive-star forming regions, with the most massive stars being the youngest in the cluster. Whether these observations really suggest an evolutionary sequence, and whether or not high-mass stars form after low-mass stars, remains a matter of debate; for a discussion see e.g. \citet{Zinnecker2008}.

VLT/X-shooter is a powerful instrument to study star formation, since both the stellar photosphere and the circumstellar material are probed with a single observation. This is also illustrated by the results of the kinematic study of HH~1042. The observed velocity structure in the jet can be reproduced reasonably well by assuming a ballistic flow with a periodic outflow rate. Incorporating relevant line excitation mechanisms in the model might improve the match between the observed intensities of the knots and the results from the simulation.

The observed jet velocity structure is consistent with the escape velocity from a launch region within a few stellar radii of an intermediate-mass star. The model parameters that best reproduce the observed emission in the position-velocity diagram are comparable with those seen in other jets \citep[e.g.][]{Velazquez2004, Raga2012}. The variations in velocity might then be caused by episodes of accretion, being of the order of the viscous timescale in the accretion disk. Another explanation would be that the observed velocity oscillation is just a projection of the global movement of the system: a binary orbit, or jet precession. 

The phase shift in velocity between the red and blue lobe would follow naturally from a precession, although the phase shift should be $\pi$ rather than $\pi/2$, which is within the uncertainty limit of the velocity of the central source. Also, the precession angle cannot be larger than $\sim 10^\circ$ due to the observed collimation of the jet. Another explanation for an asymmetry between the two lobes would be the different environmental conditions on either side of the system \citep{Melnikov2009, Podio2011}. 

\paragraph{Questions}
\textbf{R. Oudmaijer:} \textit{On what timescale can we detect proper motions of the HH~1042?}

With the limited means of performing astrometry on the SINFONI and X-shooter observations, which were taken 4 years apart, we can at best detect proper motions in the order of $1''$. These are not found. This corroborates the conclusion of the kinematics analysis, yet it does not put a firm constraint on the inclination. Based on our results we predict proper motions in the order of $0.1''$ on a timescale of $\sim 5$ yr. 

\textbf{C. Martayan:} \textit{Do you have an estimate of the inclination and spatial extent of HH~1042?}

Adopting a distance of 0.7~kpc, the projected length of the jet is $1.2 \times 10^4$~AU. Lacking proper motion measurements, we have derived the jet inclination from the disk inclination \citep[18$^\circ$, ][]{Wheelwright2010} by assuming the jet is collimated and perpendicular to the disk equatorial plane. Using this value the deprojected length is $3.9 \times 10^4$~AU. However, the adapted value for the inclination might be an underestimate due to a possible overestimated mass of the central star used by \citet{Wheelwright2010}. Note that the quoted value is the length of the optically visible part of the jet; its true extent may be much larger and visible at longer wavelengths.

\bibliographystyle{asp2010}
\bibliography{talk_Ellerbroek} 

\end{document}